\documentclass[12pt]{article}
\def\beqar {\begin{eqnarray}}
\def\eeqar {\end{eqnarray}}
\def\beq {\begin{equation}}
\def\eeq {\end{equation}}
\def\ra {{\rangle}}
\def\la {{\langle}}
\def\half {{\textstyle{1\over 2}}}
\def\Tr {{\rm Tr}}
\def\dag {{\dagger}}
\def\del {{\partial}}

\def\ep {{\epsilon}}
\def\vf {{\varphi}}
\def\bxi {{\bar \xi}}

\def\bz {\bar{z}}
\def\bZ {\bar{Z}}

\def \D {{\cal D}}

\def \H {{\cal H}}
\def \L {{\cal L}}

\def\no2 {{\textstyle{n\over 2}}}
\def \const {{\rm constant}}
\begin{document}

\begin{titlepage}
\null\vspace{-62pt}

\pagestyle{empty}
\begin{center}
\rightline{CCNY-HEP-03/4}

\vspace{1.0truein} 
{\large\bf  The effective action for 
edge states }\\
\vskip .1in
{\large\bf in higher dimensional  quantum Hall systems}\\
\vskip .3in
\vspace{.5in}DIMITRA KARABALI $^{a,c}$
 \footnote{karabali@lehman.cuny.edu}
 and V.P. NAIR $^{b,c,d}$ \footnote{vpn@sci.ccny.cuny.edu}\\
\vspace{.2in}{\it $^a$ Department of Physics and Astronomy\\
Lehman College of the CUNY\\
Bronx, NY 10468}\\
\vspace{.1in}{\it $^b$ Physics Department\\
City College of the CUNY\\  
New York, NY 10016}\\
\vspace{.1in}{\it $^c$ The Graduate School and University Center\\ 
CUNY, New York, NY 10031}\\
\vspace{.1in} {\it $^d$ Abdus Salam International Centre for
  Theoretical Physics \\
Trieste, Italy}
\end{center}
\vspace{0.5in}

\centerline{\bf Abstract}
\vskip .1in
We show that the effective action for the edge excitations of a
quantum Hall droplet of fermions in higher dimensions
is generically given by a chiral bosonic action. 
We explicitly analyze the quantum Hall effect on complex projective spaces
${\bf CP}^k$, with a $U(1)$ background magnetic field.
The edge excitations 
are described by abelian bosonic fields on $S^{2k-1}$ with
only one spatial direction along the boundary of the droplet 
relevant for the dynamics.
Our analysis also leads to an action for
edge excitations for the case of the Zhang-Hu  four dimensional
quantum
Hall effect defined on $S^4$ with an $SU(2)$ 
background magnetic field, using the fact that
${\bf CP}^3$ is an $S^2$ bundle over $S^4$.
\vskip .5in
\centerline{\it To the memory of Professor Bunji Sakita}

\baselineskip=18pt

\end{titlepage}

\hoffset=0in
\newpage
\pagestyle{plain}
\setcounter{page}{2}
\newpage

\section{Introduction}
The study of quantum Hall effect in more general
contexts than 
the classic two-dimensional plane, such as higher dimensions
and different geometries, has recently been of some research
interest, following the original analysis by Zhang and Hu
\cite{HZ}. They considered the Landau problem for charged 
fermions on $S^4$ with a background magnetic field which 
is the standard $SU(2)$ instanton. In ordinary quantum Hall effect 
a droplet of fermions occupying a certain area behaves as 
an incompressible fluid, the low energy excitations being 
area-preserving deformations which behave as massless chiral 
bosons. The motivation for considering $S^4$ was that the 
edge excitations could lead to higher spin massless 
fields, in particular the graviton, and might provide an approach to a 
quantum description of a graviton in four dimensions. In fact, in order 
to obtain a 
reasonable thermodynamic limit with a finite spatial density of particles and 
incompressibility properties, one has to consider that the fermions belong to 
a very large $SU(2)$ representation with infinite degrees of freedom. 
As a result one 
expects to find edge excitations of all values of spin present.

A number of papers have extended the original
idea of Zhang and Hu \cite{KN}- \cite{ners}. In a previous paper \cite{KN}
we studied the quantum Hall effect on even dimensional complex projective 
spaces 
${\bf CP}^k$. This is an interesting case since one can obtain incompressible 
droplet states by coupling the fermions to a $U(1)$ background field, 
thus avoiding the
infinite internal degrees of freedom. It was also suggested 
in \cite{KN} that 
the $S^4$ case with $SU(2)$ background could be understood in terms of a higher
dimensional model, ${\bf CP}^3$ with a $U(1)$ background,
using the fact that the latter is an $S^2$-bundle over $S^4$. 
This point of view has been
explored further in \cite{berneveg, zhang, ners}.

Edge excitations for droplets in $R^4$ with a $U(1)$ and $SU(2)$ background, 
essentially 
the flat limit of the Zhang-Hu model,
were studied in \cite{pol}.
Since a droplet is confined to a finite volume and
that is topologically
a neighbourhood in $R^4$,  the analysis in \cite{pol}
yields many
of the generic features
of the edge excitations. For example,
it was argued that the edge effective theory is essentially an 
infinite collection 
of one-dimensional theories and also lacks full Lorentz
invariance.

The edge excitations of a quantum Hall droplet
leads to a new class of field theories
and so, the original motivation aside, 
some further analysis is warranted. 
What would be the spectrum of edge excitations
on a general space? How is their dynamics 
related to the geometrical structures of
the space on which the droplet is?
These  are some of the questions we take up in this paper.
We find that the effective action for edge excitations
is generically a chiral action in terms of an abelian bosonic field.
Chirality here means that there is a preferred direction along the
surface of the droplet that determines the dynamics and this is chosen
by the geometry of the underlying space, where the nonrelativistic
fermions are.
The bosonic field itself can be expanded in terms of the surface oscillation
modes
of the incompressible droplet. This effectively provides a bosonic
description of the underlying nonrelativistic fermionic theory, very
similar in spirit with the collective variable approach introduced by
Sakita \cite{sakita}.

The paper is organized as follows. In section 2 we present an approach 
for deriving a generic effective 
action for edge excitations for any quantum Hall system and outline
the basic ingredients needed for such a construction. In section 3 we
analyze in detail the case of ${\bf CP}^k$ quantum Hall systems with a
$U(1)$ background field. Our analysis requires the notion of star
products and we present a simple way of writing down star products for 
${\bf{CP}}^k$ spaces. In section 4 we derive, using this method, the
well known one-dimensional chiral boson action for QHE on $S^2$.
In section 5 we apply our approach along with the
fact that ${\bf CP}^3$ can be viewed as an $S^2$-bundle over $S^4$ to
derive the effective edge action for the four dimensional quantum Hall
system introduced by Zhang and Hu \cite{HZ}. We conclude with a
summary of our results.

\section{The effective action for edge states}

We will begin with a general discussion of the effective action
for edge states for any quantum droplet. The strategy we follow is an
adaptation of a method used by Sakita \cite{sakita} to derive the edge 
effective
action in the case of the conventional two-dimensional quantum Hall effect.
The basic 
observation is very simple. The choice of the droplet 
we are considering is specified quantum mechanically 
by a density matrix ${\hat \rho}_0$. Once this state 
is chosen, time-evolution being unitary, any density 
matrix which is connected to ${\hat \rho}_0$ by time-evolution 
is of the form ${\hat\rho} = {\hat U} {\hat \rho}_0 {\hat U}^\dagger$,
where ${\hat U}$ is a unitary operator. The entire dynamical 
information is in ${\hat U}$ and we can write down an action
for ${\hat U}$ as
\beq
S= \int dt~ \left[ i \Tr ({\hat \rho}_0 { \hat U}^\dagger \del_t {\hat U})
~-~ \Tr ({\hat \rho}_0 {\hat U}^\dagger {\cal H} {\hat U}) \right]
\label{1}
\eeq
where $\H$ is the Hamiltonian.
Variation of ${\hat U}$ leads to the extremization condition for
$S$ as
\beq
i {\del {\hat \rho}\over \del t}
= [ \H , {\hat \rho}]
\label{2}
\eeq
which is the expected evolution  equation for the density
matrix. The strategy for obtaining an effective action 
for edge states is to evaluate various quantities in 
(\ref{1}) as classical functions in a suitable 
`semiclassical' approximation, in the limit of a large 
number of available states.

If we apply this in full generality to a quantum Hall 
droplet with $K$ fermions, ${\hat \rho}_0$ should be 
specified as $\vert \alpha \ra \la \alpha \vert$,
where $\vert \alpha\ra$ is a state in the full Fock space
which is of the form
\beq
\vert \alpha \ra = \int_{x_i}~f (x_1, x_2, ...,x_K)~ 
{\psi^\dag}(x_1) {\psi^\dag}(x_2)...{\psi^\dag}(x_K)
~\vert 0\ra \label{3}
\eeq
where $\psi^\dag$'s are creation operators for fermions and
$f(x_1, x_2, ..., x_K)$ is the $K$-body wavefunction
of the droplet state. The function $f (x_1, x_2, ..., x_K)$
can include the effects of many-particle correlations.
In practice, for a $\nu =1$ droplet, one can take
$f(x_1, x_2, ..., x_K)$ to be a product of effective
one-particle functions suitably antisymmetrized,
in other words, a Hartree-Fock approximation.
If we further neglect interparticle interactions, concentrating
on the edge dynamics, $\H$ is given in the form
$\int \psi^\dag {\hat h} \psi$ and the traces and
operators in (\ref{1}) reduce to the one-particle
Hilbert space. In this case, ${\hat \rho}_0 =
\sum_{i=0}^K  \vert i\ra \la i\vert$ where $\vert i\ra$'s
are the occupied one-particle states and ${\hat U}$ is 
now an operator on this one-particle Hilbert space. 
Since we will be applying this to quantum Hall droplets on 
compact spaces like ${\bf CP}^k$, we take the 
one-particle Hilbert space to have a finite dimension 
$N$, so that ${ \hat U}$ is a unitary $(N\times N)$-matrix.
$K$ states in this Hilbert space are occupied
and we will simplify all the expressions in the limit
$N\rightarrow \infty$, and $K$ very large, but finite.
For an abelian magnetic field, the large
$N$-limit of ${\hat U}$ becomes an abelian field characterizing the
surface deformations of the droplet.

There are basically two ingredients needed to extract 
an edge state action
from (\ref{1}).
\begin{enumerate}
\item We need ${\hat \rho}_0$ , in the large $N$ limit,
to be constant over the phase volume occupied by the
droplet $\D$, so that its derivative tends to a $\delta$-function
with support at the edge of the droplet.
\item Commutators between operators tend to suitable
Poisson brackets as $N\rightarrow \infty$ (which is not
the usual $\hbar \rightarrow 0$ limit).
\end{enumerate}
For the ${\bf CP}^k$ cases of interest, these results
will be shown in the next section. The states will be characterized
by an integer $n\sim BR^2$ where
$B$ is the magnetic field and $R$ is the radius of the
${\bf CP}^k$; the dimension $N\sim n^k$ and scaling is
done with $R\sim \sqrt{n}$.
With this preparation,
the effective action is easily obtained. 
\beqar
i \Tr ( {\hat \rho}_0 { \hat U}^\dag \del_t {\hat U})&=&
i \Tr \left[  {\hat \rho}_0 e^{-i{ \hat \Phi}} {\del \over \del t}
e^{i{ \hat \Phi}}\right]\nonumber\\
&=& - \int_0^1 d\alpha~ \Tr \left[ {\hat \rho}_0 
e^{-i\alpha { \hat \Phi}} (\del_t {\hat \Phi} ) e^{i\alpha { \hat \Phi}}
\right] \nonumber\\
&=& -\int_0^1 d\alpha~ \Tr \left[ e^{i\alpha {\hat \Phi}} {\hat \rho}_0
e^{-i\alpha { \hat \Phi}} ~\del_t { \hat \Phi} \right]\nonumber\\
&=& -\int_0^1 d\alpha ~ \Tr \left[ \left( {\hat \rho}_0 
+i\alpha [{ \hat \Phi} ,{\hat \rho}_0 ]
+{i^2 \alpha^2 \over 2} [{\hat \Phi} ,[{\hat \Phi} ,{\hat \rho}_0]]
+\cdots \right)~\del_t {\hat \Phi}\right] \nonumber\\
\label{4}
\eeqar
As $N\rightarrow \infty$,
\beqar
[ {\hat \Phi} , {\hat \rho}_0 ] &\approx& i{c\over n} ~\{ \Phi , \rho_0\}
\nonumber\\
{}[ {\hat \Phi} ,[{\hat \Phi} , {\hat \rho}_0 ] ] &\approx& i^2 
\left({c\over n}\right)^2 
\{ \Phi , \{ \Phi , \rho_0 \}\} , \hskip .2in {\rm etc.,}
\label{5}
\eeqar
where $c$ is a constant of order one and $\Phi ,\rho_0$
are the classical functions corresponding to ${\hat \Phi} , {\hat \rho}_0$. The Poisson
brackets in (\ref{5}) are calculated using rescaled, dimensionless coordinates. Further,
\beq
\Tr \approx N \int d\mu \label{6}
\eeq
where $d\mu$ is the volume element on the phase space.
The large $N$ limit of (\ref{4}) is thus
\beq
i \Tr ({\hat \rho}_0 U^\dag \del_t U ) \approx
{Nc \over 2n}\int d\mu~ \{ \Phi , \rho_0\}~ \del_t \Phi
\label{7}
\eeq
where we have dropped terms 
which are total time-derivative.

Let $\Omega_{ij}$ be the symplectic two-form of the phase space
corresponding to the Hilbert space of one-particle states.
Then
\beqar
\{ \Phi , \rho_0 \}&=& (\Omega^{-1}) ^{ij} {\del \Phi \over
\del \xi^i} {\del \rho_0 \over \del \xi^j}\nonumber\\
&=& (\Omega^{-1})^{ij} {\hat e}_j {\del \Phi \over \del \xi^i}
{\del \rho_0 \over \del r}\nonumber\\
&=& (\L\Phi ) \left( {\del \rho_0 \over \del r^2}\right)
\label{8}
\eeqar
where
\beq
\L\Phi = \left[ 2r {\hat e}_j (\Omega^{-1})^{ij}
{\del \Phi \over \del \xi^i}\right]
\label{8a}
\eeq 
The coordinates $\xi_i$ are dimensionless, measured in units of the radius $R$ of the compact
space, $\xi_i = (x_i + iy_i) / R$. 
${\hat e}_j$ is a unit vector normal to the boundary $\del\D$ of the
droplet and $r=\sqrt{\sum \xi_i \xi_i}$ is the dimensionless, normal coordinate. $\L\Phi$
involves only derivatives of $\Phi$ with respect to a tangential direction. 
In deriving (\ref{8})
we have used the fact that $\rho _0$ depends only on the normal coordinate r.
Thus
\beqar
i \Tr ({\hat \rho}_0 U^\dag \del_t U)&\approx&
{Nc \over 2n}\int d\mu~ {\del \rho_0 \over \del r^2}~\left[
{\del \Phi \over \del t} ~\L\Phi\right]
\label{9}
\eeqar

The term involving the Hamiltonian can be simplified as follows.
For the Landau level under consideration, we have $N$ degenerate
states of energy $E_0$ to begin with, the degeneracy being lifted
by a potential ${\hat V}$; thus $\H = E_0 +{\hat V}$.
The constant term $E_0$
is not of interest for edge dynamics, so we can write
\beqar
\Tr ({\hat \rho}_0 U^\dag \H U)&=& \const + \int_0^1 d\alpha~
\Tr ( {\hat \rho}_0 e^{-i\alpha {\hat \Phi}}
i [{\hat V},{\hat \Phi} ] e^{i\alpha {\hat \Phi}} )\nonumber\\
&=& \const + \Tr \left( {\hat \rho}_0 i [{\hat V}, {\hat \Phi} ] + {1\over 2}
[ {\hat \rho}_0 , {\hat \Phi} ] [{\hat V}, {\hat \Phi} ] +\cdots \right)
\nonumber\\
&=& \const +i \Tr ([{\hat \rho}_0 , {\hat V}] {\hat \Phi} )
+{1\over 2}\Tr ( [{\hat \rho}_0 , {\hat \Phi} ] [ {\hat V}, {\hat \Phi} ])+
\cdots
\nonumber\\
&\approx& \const + {Nc\over n} \int d\mu ~\{ V, \rho_0 \} \Phi ~-~
{Nc^2 \over 2n^2} \int d\mu ~\{ \rho_0 , \Phi \} \{  V, \Phi \}
+\cdots \nonumber\\
\label{10}
\eeqar
We choose ${\rho}_0$ to minimize the potential energy, the particles
being located at states of minimum available energy.
$V$ is thus a constant along directions tangential
to the boundary of the droplet $\del \D$, it depends only on the 
normal coordinate $r$. Further
$\{ V, \rho_0\}=0$.
The last term in (\ref{10}) simplifies as
\beq
\{\rho_0 ,\Phi \} \{ V, \Phi \} =
{\del \rho_0 \over \del r^2}~ (\L\Phi )~
~{\del V \over \del r^2}~(\L\Phi ) 
\label{11}
\eeq
Putting all this together, we have, for the
effective action for the
edge states,
\beq
S \approx {Nc \over 2n}\int dtd\mu \left( {\del \rho_0 \over \del r^2}\right)~
\left[ {\del \Phi \over \del t} (\L\Phi ) 
+{c\over n} {\del V\over \del r^2} (\L\Phi )^2
\right] + \const +{\cal O}(1/n)
\label{12}
\eeq
where the operator $\L$ in the above expression is the $n \rightarrow \infty$
limit of the expression (\ref{8a}).
\beq
\L\Phi = \left[ 2r {\hat e}_j (\Omega^{-1})^{ij}
{\del \Phi \over \del \xi^i}\right]_{n \rightarrow \infty}
\label{12c}
\eeq 
As we mentioned earlier, in the large $N$ limit, 
$\del \rho_0 /\del r^2$ is a $\delta$-function with support only at
the edge of the droplet; as a result the $r$-integration
can be done and the action is entirely on the boundary of
the droplet. Since the radius of the space
scales like $\sqrt{n}$, the integral will give an extra factor
of $n^{-k+1}$ for a compact $2k$ dimensional space. For the 
${\bf CP}^k$ cases we analyze in detail in the next section, we show that 
$N \sim n^k$, so that the limit $n\rightarrow \infty$ gives a finite
prefactor to the action.
Also the potentials we choose will be such that
$\del V/ \del r^2 \sim n$ and so, $\omega = {c \over n} (\del V/\del r^2)$
is finite as $n\rightarrow \infty$. Notice that if the potential
does not have this property, then $\omega$ goes to zero
or infinity as $n\rightarrow \infty$. In the first case,
there are zero energy modes for the field $\Phi$,
indicating that the potential has not uniquely confined
the droplet to the shape initially chosen. In
the second case, no fluctuations of the droplet are energetically
allowed. It is thus physically reasonable to
take the potential to scale as $n$ leading to a
constant nonzero $\omega$ as $n\rightarrow \infty$. The final effective 
edge action 
(up to a multiplicative and additive constant) is of 
the form
\beq
S \approx - \int_{\del\D} dt 
\left[ {\del \Phi \over \del t} (\L\Phi ) 
+\omega (\L\Phi )^2
\right] 
\label{12b}
\eeq

We see from (\ref{12b}) that the effective edge dynamics 
involves only the time-derivative of $\Phi$ and
one tangential derivative given by $\L\Phi$;
the action is in
this sense chiral.
The field $\Phi$ can, and does indeed, depend on the 
remaining tangential
directions,
which we will denote by $\Sigma$, leading to a
multiplicity of modes.
We can expect the 
multiplicity to be parametrized by $\Sigma$, but this
is not quite accurate.
For a spherical droplet in even dimensions $2k$,
$\del\D \sim S^{2k-1}$ and therefore the space
$\Sigma$ 
is ${\bf CP}^{k-1}$. However, $S^{2k-1}$ is not 
the product ${\bf CP}^{k-1}\times S^1$ and so
$\Phi$ cannot be expanded in terms of functions
on ${\bf CP}^{k-1}$, but rather will involve
sections of vector bundles on ${\bf CP}^{k-1}$.
This will become clear from the examples in later 
sections.

\section{Analysis on ${\bf CP}^k$}

\noindent{$\underline{Landau ~levels ~and ~wavefunctions}$}
\vskip .1in
In the last section we derived the general form of the effective action
for the edge states. To complete this derivation, we need to 
explicitly show the
emergence of the Poisson brackets and also that the density $\rho_0$ is 
constant over
the droplet $\D$. We will do this now
for ${\bf CP}^k$ with a $U(1)$ background magnetic field. 
Our approach follows the group theoretic analysis used in our previous paper 
\cite{KN}.
Since
${\bf CP}^k ={SU(k+1) / U(k)}$,
the required wavefunctions can be obtained
as functions on 
$SU(k+1)$ which have a specific charge under the $U(1)$ subgroup. 
Let $t_a$ denote the generators of $SU(k+1)$
as matrices in the fundamental representation; we normalize them
by $\Tr (t_a t_b )=\half \delta_{ab}$.
The generator corresponding to the $U(1)$ direction
of the subgroup $U(k)$ will be denoted by 
$t_{k^2+2k}$. 
A basis of functions on $SU(k+1)$
is given by the Wigner $\cal{D}$-functions which are the
matrices corresponding to the
group elements in a representation
$J$
\beq
\D^{(J)}_{L;R}(g) = \la J ,L_i \vert {\hat g}\vert J, R_i \ra \label {13a}
\eeq
where $L_i,~R_i$ stand for two sets of quantum numbers specifying the 
eigenvalues of the
diagonal generators for left and right $U(k)$ actions on $g$, respectively.
On an element $g\in SU(k+1)$, considered as a $(k+1) \times (k+1)$-matrix,
we can define left and right $SU(k+1)$ actions by
\beq
{\hat{L}}_a ~g = t_a ~g, \hskip 1in {\hat{R}}_a~ g = g~t_a
\label{14}
\eeq
As discussed in \cite{KN}, the wavefunctions for the Landau 
levels on ${\bf CP}^k$ with a $U(1)$ background field are singlets under 
the subgroup 
$SU(k)_R$ and carry $U(1)_R$ charge as specified by the background field. 
This implies 
that the quantum numbers $R_i$ in (\ref{13a}) are constrained to be
\beqar
&&R_j =0 ~~~~~~~~~~~~~~~j=1,\cdots , k^2 -1 \nonumber \\
&&R_{k^2 + 2k} = -nk {1 \over {\sqrt{2k(k+1)}}}
\label{14a}
\eeqar
$n$ goes like $BR^2$ where $B$ is the magnetic field and
$R$ is the radius of the ${\bf CP}^k$; $n$, upto factors involving
the normalization of generators, is an integer in accordance
with the Dirac quantization rule. 

There are $2k$ right generators of $SU(k+1)$ which are not in
$U(k)$; these can be separated into $t_{+i}$ which are
of the raising type and $t_{-i}$ which are of the lowering
type. The covariant derivatives on ${\bf CP}^k$ 
can be identified with these ${\hat R}_{\pm i}$ right rotations on $g$. 
This is
consistent with the fact that the commutator of
covariant derivatives
is the magnetic field. 
Since the Hamiltonian goes like
${\hat R}_{+i} {\hat R}_{-i}$, the lowest Landau level will
also obey the condition ${\hat R}_{-i} \Phi =0$.
One can therefore identify the lowest Landau level
(LLL) states as the lowest weight vector for the 
representation corresponding to the right action.

The Landau levels correspond to the various possible irreducible 
$SU(k+1)$ 
representations $J$
which contain an $SU(k)$ singlet. Such representations can be labelled by two 
integers 
$J = (p, q)$ such that $p-q = n$. The lowest Landau level corresponds to 
$p =n,~~q=0$. These are completely symmetric representations.
The dimension of this representation $J = (n, 0)$ is
\beq
dim J = {{(n+k)!} \over {n! k!}} \equiv N
\label{14b}
\eeq
$N$ expresses the number of states in the lowest Landau level.

The LLL wavefunctions on ${\bf CP}^k$ can be written as
\beq
\Psi^{(n)}_m (g) = \sqrt{N}~\D^{(n)}_{m; -n} (g) \label{15}
\eeq
We denote the fixed state for the right action on the Wigner 
function above as $-n$, indicating that the eigenvalue for 
${\hat R}_{k^2 + 2k}$ is $-nk / \sqrt{2k(k+1)}$ as in (\ref{14a}). 
The index $m$  
specifies the state within the lowest Landau level. 
 
The wavefunctions (\ref{15})
are normalized by virtue of the orthogonality theorem
\beq
\int d\mu (g) ~\D^{*(n)}_{m;-n} (g)~\D^{(n)}_{m';-n}
(g) ~=~ {\delta_{mm'}\over N}
\label{15a}
\eeq
where $d\mu (g)$ is the Haar measure on $SU(k+1)/U(k)$, normalized to
unity ($d\mu (g)$ can be taken as the Haar measure on the full group
$SU(k+1)$, properly normalized, since the integrand is $U(k)$ 
invariant.)

Since the LLL wavefunctions correspond to completely symmetric 
representations, 
a convenient way to represent them is in terms of the group elements
\beq
g_{\alpha, k+1}\equiv u_{\alpha}~~~~~~~\alpha=1, \cdots , k
\label{15b}
\eeq
In terms of $u_{\alpha}$'s, the corresponding Wigner $\D$ functions are of 
the form
$\D \sim u_{\alpha_1} u_{\alpha_2}\cdots u_{\alpha_n}$. Eq. (\ref{15b}) 
also implies that $u^*\cdot u =1$
and so one may parametrize them as
\beqar
u_i &=& {\xi_i \over \sqrt{1+\bxi \cdot \xi }}~~~~~~~~~~~~~i=1,\cdots , k 
\nonumber \\
u_{k+1} &=& {1\over \sqrt{1+\bxi \cdot \xi }}
\label{16}
\eeqar
$\xi, ~\bxi$ can be thought of as the complex local coordinates for 
${\bf CP}^k$.
With this parametrization, the wavefunctions for the lowest Landau
level are
\beqar
\Psi^{(n)}_m &=& \sqrt{N}~ \D^{(n)}_{m, -n}(g) \nonumber\\
\D^{(n)}_{m, -n}(g) &=& \left[ {n! \over i_1! i_2! ...i_k!
(n-s)!}\right]^\half ~ {\xi_1^{i_1} \xi_2^{i_2}\cdots \xi_k^{i_k}\over
(1+\bxi \cdot \xi )^{n \over 2}} \nonumber\\
s &=& i_1 +i_2 + \cdots +i_k \label{17}
\eeqar
The condition ${\hat R}_{-i} \D^{(n)}_{m,-n}=0$ is a holomorphicity
condition and this is reflected in the fact that the
wavefunctions for the lowest Landau level are holomorphic
in $\xi$'s, apart from certain overall factors.
The states (\ref{17}) are coherent states for ${\bf CP}^k$.
The inner product for the $\Psi$'s may be written in these coordinates
as
\beqar
\la \Psi \vert \Psi' \ra &=& \int d\mu ~\Psi^* ~\Psi'\nonumber\\
d\mu &=& {k!\over \pi^k}{d^k \xi d^k\bxi \over (1+\bxi \cdot \xi )^{k+1}}
\label{18}
\eeqar

We will also need the formulae for the left and right translations (\ref{14})
as differential operators. 
The left rotations ${\hat L}_a~ a=1, \cdots , k^2 + 2k$ correspond to
magnetic translations. One 
can easily write down the corresponding 
differential operator in terms of $u_{\alpha}$'s.
\beq
{\hat L}_a = u_{\beta} (t^a)_{\alpha \beta} {\del \over {\del u_{\alpha}}} -
u^*_{\alpha} (t^a)_{\alpha \beta} {\del \over {\del u^*_{\beta}}}
\label{18a}
\eeq
The right generators cannot be easily written in terms of $u_{\alpha}$'s; 
we express them in 
terms of the group parameters $\vf^i$
\beq
{\hat R}_a = i (E^{-1})^i_a {\del \over \del \vf^i}
\label{19}
\eeq
where 
\beq
g^{-1} dg = -i t_a E^a_i ~d\vf^i
\label{20}
\eeq

\vskip .1in\noindent
{$\underline{Star ~products, ~commutators ~and ~Poisson ~brackets}$}
\vskip .1in

Let $\hat{A}$ be a general matrix acting on the $N$-dimensional
Hilbert space of lowest Landau level states with matrix elements $A_{ms}$.
We define the symbol corresponding to $A$ as the function \footnote{  
The symbol as defined in (\ref{21}) is related to the expectation value
of $A$ in the coherent state basis; it coincides with the expectation value
in the large $n$ limit.}
\beq
A(g) = A(\xi ,\bxi )= \sum_{ms}\D^{(n)}_{m, -n}
(g) A_{ms} \D^{*(n)}_{s, -n}(g)
\label{21}
\eeq

We are interested in the 
symbol corresponding to the product of two matrices
$A$ and $B$.
This can be written as
\beqar
(AB)(g)&=& \sum_r A_{mr} B_{rs} \D^{(n)}_{m, -n}
(g) \D^{*(n)}_{s, -n}(g) \nonumber\\
&=& \sum_{rr'p} \D^{(n)}_{m, -n}(g) A_{mr}
~\D^{*(n)}_{r,p}(g) \D^{(n)}_{r',p}(g)~
B_{r's} \D^{*(n)}_{s, -n}(g)
\label{22}
\eeqar 
using $\delta_{rr'}=\sum_p
\D^{*(n)}_{r,p}(g) 
\D^{(n)}_{r',p}(g)$.
The term with $p =- n$ on the right hand side
of (\ref{22}) gives the product of the symbols for $A$ and $B$.
The terms with $p > -n$ may be written using raising
operators as
\beq
\D^{(n)}_{r,p}(g) = \left[{(n-s)!\over n! i_1! i_2! \cdots i_k!} 
\right]^\half {\hat R}_{+1}^{i_1}~{\hat R}_{+2}^{i_2} \cdots
{\hat R}_{+k}^{i_k} ~\D^{(n)}_{r,-n}(g)
\label{23}
\eeq
Here $s=i_1+i_2+\cdots +i_k$
and the $t_{k^2+2k}$-eigenvalue for the state
$\vert p\ra$ is $(-nk +sk +s )/\sqrt{2k(k+1)}$.
Since ${\hat R}_{+i} \D^{*(n)}_{s, -n} =0$,
we can also write
\beq
\biggl[{\hat R}_{+i}\D^{(n)}_{r',-n}(g)\biggr] B_{r's} \D^{*(n)}_{s, -n} (g)
= \biggl[{\hat R}_{+i} 
\D^{(n)}_{r',-n}
B_{r's} \D^{*(n)}_{s, -n} (g)\biggr] = {\hat R}_{+i} B(g) 
\label{24}
\eeq
Further keeping in mind that ${\hat R}_+^* = - {\hat R}_-$,
we can combine (\ref{22}-\ref{24}) to get
\beqar
(AB)(g) &=& \sum_s (-1)^s \left[ {(n-s)! \over n! s!}\right]
\sum_{i_1+i_2+\cdots +i_k=s}^n~
{s! \over i_1! i_2! \cdots i_k!}~{\hat R}_{-1}^{i_1} {\hat R}_{-2}^{i_2}
\cdots {\hat R}_{-k}^{i_k} A(g)\nonumber\\
&&\hskip 2.5in\times~
{\hat R}_{+1}^{i_1} {\hat R}_{+2}^{i_2}\cdots {\hat R}_{+k}^{i_k}
 B(g)\nonumber\\
&\equiv& A(g) * B(g)
\label{25}
\eeqar
The right hand side of this equation is what is known as the
star product for functions on ${\bf CP}^k$. It has been written down
in different forms in the context of noncommutative 
${\bf CP}^k$ spaces \cite{bal}; our argument here
follows the presentation in \cite{nair} which gives a simple and general way of
constructing star products.
The first term of the sum on the right hand side gives $A(g) B(g)$,
successive terms involve derivatives and are down by powers of
$n$, as $n\rightarrow \infty$. For the symbol corresponding 
to the commutator of $A, ~B$, we have
\beq
\bigl( [A,B]\bigr)(g) = -{1\over n}
({\hat R}_{-i} A ~{\hat R}_{+i} B - {\hat R}_{-i} B~ {\hat R}_{+i} A )
~+~ {\cal O} (1/n^2)\label{26}
\eeq

The K\"ahler two-form on ${\bf CP}^k$
may be written as
\footnote{{The symplectic form which via geometric quantization
gives the states we are considering is proportional to
the K\"ahler form.}}
\beqar
\Omega &=&-i \sqrt{2k \over k+1}
\Tr \bigl( t_{k^2+2k} ~g^{-1}dg\wedge g^{-1}dg \bigr)\nonumber\\
&=& -{1\over 4} \sum_{i=1}^k~\ep_{a_i b_i} E^{a_i}_c E^{b_i}_d 
~d\vf^c \wedge d\vf^d
\nonumber\\ 
&\equiv& {1\over 2} \Omega_{cd} ~ d\vf^c \wedge d\vf^d
\label{27}
\eeqar
Functions $A,~B$ on ${\bf CP}^k$ obey the condition 
${\hat R}_\alpha A = {\hat R}_\alpha B =0$, where ${\hat R}_\alpha$
(with the index $\alpha$)
is any generator of the subgroup $U(k)$.
With this condition, we find
\beq
i \sum_i \ep^{a_i b_i} (\hat{R}_{a_i} A) (E^{-1})^c_{b_i} ~\Omega_{cd}
= - {\del A \over \del \vf^d}
\label{28}
\eeq
This shows that $i\sum_i \ep^{a_i b_i}(\hat{R}_{a_i} A)
(E^{-1})^c_{b_i}$ 
may be
identified as the Hamiltonian vector field
corresponding to $A$. The Poisson bracket of $A, ~B$, as defined
by $\Omega$, is
thus given by
\beqar
\{ A, B\} &=& (\Omega^{-1})^{ab} 
{\del A \over {\del \vf^a}} {\del B \over {\del \vf^b}} 
\nonumber\\
&=& i \sum_{i=1}^{k}\left( \hat{R}_{-i} A~ \hat{R}_{+i} B ~-~ 
\hat{R}_{-i}B ~\hat{R}_{+i}A\right)
\label{29}
\eeqar
Combining with (\ref{26}), we find
\beq
\left( [A,B]\right) (g) = {i\over n} \{ A, B\} ~+~ {\cal O}(1/n^2)
\label{30}
\eeq
This completes the demonstration of
equation (\ref{5}) with $c =1$ for ${\bf CP}^k$.
If desired, one can also write the Poisson bracket in terms of the local
coordinates $\xi ,\bxi$ introduced in (\ref{16}).
The relevant expressions are
\beqar
\Omega &=& -i \left[ {d\bxi \cdot d\xi
\over (1+\bxi \cdot \xi ) }- {\xi \cdot d\bxi ~\bxi\cdot d\xi \over
(1+\bxi \cdot \xi )^2}\right]\nonumber\\ 
\{ A, B\}&=& i (1+\bxi \cdot \xi ) 
\Biggl( {\del A \over \del \xi^i}
{\del B\over \del \bxi^i} - {\del A \over \del \bxi^i }{\del B\over\del \xi^i}
 \nonumber\\
&&\hskip .2in+\xi\cdot {\del A\over \del\xi} 
\bxi \cdot {\del B\over\del\bxi} -\bxi\cdot {\del A \over
  \del\bxi}
\xi\cdot {\del B\over \del\xi} \Biggr)
\label{31}
\eeqar 

The trace of an operator ${\hat A}$ may be written as
\beqar
\Tr {\hat A} &=& \sum_m {A}_{mm} = N~\int d\mu (g) 
\D^{(n)}_{m, -n} ~A_{mm'}~ 
\D^{*(n)}_{m', -n}\nonumber\\
&=& N \int d\mu (g)~ A(g)
\label{32}
\eeqar
which is consistent with (\ref{6}). The trace of the product
of two operators $A, B$ is given by
\beq
\Tr {\hat A} {\hat B} ~=~ N \int d\mu (g) ~A(g)*B(g)
\label{33}
\eeq
\vskip .1in\noindent
{$\underline{The ~density ~matrix}$}
\vskip .1in
We now consider the density $\rho_0$ as we fill up states
and make a droplet.  If we decompose the
representation of $SU(k+1)$ in terms of $SU(k)$ representations,
we find that it is made up of the
singlet of $SU(k)$, the fundamental of $SU(k)$, the rank
two symmetric representation, etc., each occuring once.
There are many ways to fill a part of these states, giving
different ${\hat \rho}_0$'s; the results for the edge
states will be different but qualitatively similar.
The potentials break translation symmetry and are
functions of ${\hat L}_a$ in general. One can choose
the potential appropriately to produce a chosen ${\hat \rho}_0$.
The simplest case of a droplet
would be to fill up a certain number of
complete $SU(k)$ representations, preserving an $SU(k)$ symmetry.
The potential which does this can be taken as a
function of ${\hat L}_{k^2+2k}$.
For example, we can take
\beq
{\hat V}= \sqrt{2k \over k+1}~\omega \left( {\hat L}_{k^2+2k} + 
{nk\over \sqrt{2k(k+1)}}\right)
\label{33a}
\eeq
The normalization above has been chosen so that if $s$ is the rank of 
the symmetric representation $SU(k)$ then
\beq
\la s | \hat{V} | s \ra = \omega s
\label{33b}
\eeq
This will fill up states from $m= -n$ to
the complete symmetric representation of $SU(k)$ of, say,
rank $M$.
This corresponds to ${\hat\rho}_0 =
\sum_{s=0}^M \sum_\lambda \vert s\lambda\ra \la s\lambda\vert$
where $\lambda$ designates states within the rank $s$
representation of $SU(k)$.
This ${\hat \rho}_0$ is
the identity for the first $M$ $SU(k)$ representations
and zero for the remainder. 
As a result, the commutator of
$\hat{L}_a$ will have contributions only from states near the
last rank-$M$ $SU(k)$ representation, i.e., near
the edge. More specifically, the symbol for the density matrix
for this choice is
\beqar
\rho_0&=& \sum_{s=0}^M \sum_\lambda \D^{(n)}_{s\lambda ,-n}
\D^{*(n)}_{s\lambda , -n}\nonumber\\
&=& \sum_{s=0}^M \left[ {n! \over s! (n-s)!}\right]
{({\bxi\cdot\xi})^s \over (1+\bxi\cdot\xi )^n}\label{34}
\eeqar 
Taking the derivative of $\rho_0 (\xi, \bxi)$ with respect to 
$ \bxi \cdot \xi$ we find
\beq
{\del \rho_0 \over {\del ( \bxi \cdot \xi )}} = -{n! \over {(1+ \bxi
    \cdot \xi)^n}} 
{{(\bxi \cdot \xi)^M} \over {(n-M-1)! M!}}
\label{34a}
\eeq
The coordinates $\xi ,\bxi$ are in units of the 
radius ${\bf CP}^k$, namely $\xi = (x+iy)/R$. Since $n \sim BR^2$
we define $v = n\bxi \cdot \xi$ where $v$ is $n$ independent, 
and take the $n\rightarrow
\infty$ of (\ref{34a}). Using the fact that 
$(1 + {v \over n})^n \rightarrow e^v$ as $n \rightarrow \infty$ and the 
Stirling's formula
\beq
n! \sim \sqrt{2 \pi} e^{-n} n^{n+\half}
\label{34b}
\eeq
we find
\beq
{\del \rho_0 \over \del v} \approx -   e^{-v} {v^M\over M!}
\label{35}
\eeq
For large $M$, (large number of fermions) the 
derivative of $\rho_0$ is sharply peaked at $v\approx M$. Expanding around 
$M$ and using Stirling's formula again, we find \cite{sakita}
\beqar
{\del \rho_0 \over \del v} &\approx& - {e^{M-v} v^M \over 
{M^M \sqrt{2\pi M}}} \approx - {1\over \sqrt{2\pi M}}
\exp\left( - \half M (1-v/M)^2\right)\nonumber\\
&\approx& - {1\over M} \delta\left( 1-{v\over M}\right),
\label{36}
\eeqar
as $M$ becomes very large. Thus
\beq
\rho_0 \approx \Theta\left( 1-{v\over M}\right)
= \Theta \left( 1 - {n\bxi \cdot \xi \over M}\right)
\label{37}
\eeq
The radius of the droplet will be proportional to $\sqrt{M}$, in fact $r_D^2 = M/B$, where $B$
is the magnetic field. We have thus shown the 
required property of $\rho_0$.
\vskip .1in\noindent
{$\underline{Calculation ~of ~the ~potential ~energy}$}
\vskip .1in

Finally we obtain the large $n$ limit of the potential energy
$Tr(\hat{\rho} _0 \hat{U} ^{\dagger} \hat{V} \hat{U})$ corresponding 
to the $SU(k)$ 
symmetric potential chosen in (\ref{33a}). In order to do this we have to 
calculate the classical function or symbol corresponding to $\hat{V}$ as 
outlined in (\ref{10}). Using our definition (\ref{21}) we get
\beq
V= \omega \sqrt {2k \over {k+1}} \left[ \la -n | g^{\dagger}~ 
\hat{L} _{k^2 + 2k}~ g | -n \ra + {nk \over {\sqrt{2k(k+1)}}} \right]
\label{38}
\eeq
Using
\beqar
&& g^{\dagger} \hat{L} _a g = S_{ab} \hat{L} _b \nonumber \\
&& S_{ab} = 2 Tr (g^{\dagger} t_a g t_b)
\label{38a}
\eeqar
we find that
\beqar
V &=& \omega \sqrt {2k \over {k+1}} \left[ \la -n | L_{k^2 + 2k} 
 | -n \ra ~S_{k^2 +2k, k^2 + 2k} + {nk \over {\sqrt{2k(k+1)}}} \right] 
\nonumber \\
& = & \sqrt{2k \over k+1}\omega \left( - {k n \over \sqrt{2k(k+1)}}
S_{k^2+2k,k^2+2k} +{nk\over\sqrt{2k(k+1)}}\right)
\label{38b}
\eeqar
A straightforward calculation gives
\beq
S_{k^2 +2k, k^2 + 2k} = {{k - \bxi \cdot \xi} \over {1 + \bxi \cdot \xi}}
\label {38c}
\eeq
Combining (\ref{38a}), (\ref{38b}) we find
\beq
V = \omega n {{\bxi \cdot \xi} \over {1 + \bxi \cdot \xi}}
\label {38d}
\eeq
Notice that ${{\del V} \over \del {(\bxi \cdot \xi)}} \sim n$ as indicated in section 2.
Further as $n \rightarrow \infty$, $V$ is just a harmonic oscillator confining potential.

In a similar fashion one can calculate the classical function corresponding to
any potential which is an arbitrary function of $\hat{L}_a$'s.

\vskip .1in\noindent
{$\underline{The ~effective ~action ~and ~nature ~of ~edge ~states}$}
\vskip .1in
The formula (\ref{37}) for the density shows that the boundary of
the droplet is defined by $n \bxi\cdot \xi =M$; the radius 
of the droplet is $\sqrt{M/B}$.
The boundary thus looks like a sphere
$S^{2k-1}$. It is then convenient to go over to
angular coordinates to evaluate the effective action.
Changing to $\sqrt{n}\xi$ as the coordinate, we get a factor
of $n^{-k}$ from the volume measure. Recalling that
$N= (n+k)! /n!k!\sim n^k /k!$, we find from (\ref{12}) 
\beq
S_{{\bf CP}^k}\approx -{1\over 4\pi^k} M^{k-1}\int d\Omega_{S^{2k-1}}
 \left[ {\del\Phi
\over \del t} (\L\Phi ) + \omega
\left( \L\Phi \right)^2 \right]
\label{42}
\eeq
where $d\Omega_{S^{2k-1}}$ denotes the volume element on the sphere
$S^{2k-1}$; the factor $M^{k-1}$ is as expected for
a droplet of radius $\sim \sqrt{M}$.
The operator $\L$ is identified in terms of the coordinates
$\bxi ,\xi$ as
\beq
\L = i \left( \xi\cdot {\del \over \del \xi} -
\bxi\cdot {\del \over \del \bxi}\right)
\label{43}
\eeq
This operator is the one defined in (\ref{8a}); terms which 
vanish as 
$n \rightarrow \infty $ have been dropped.

We will now consider the nature of the edge excitations.
We can expand $\Phi$ in powers of $z_i =\sqrt{n}~\xi_i$ and
$\bz_i = \sqrt{n}~\bxi_i$.
Since $\bz\cdot z$ is fixed at the droplet boundary, $\Phi$ 
is obviously a function on
$S^{2k-1}$. We can write $z =h z_0$, where
$z_0 =(0,0,...,0,\sqrt{M})$, $h\in SU(k)$.
This naturally leads to $SU(k)/SU(k-1)= S^{2k-1}$.
Powers of $z,\bz$ are of the form
$\bz^{i_1} \bz^{i_2}...\bz^{i_q} z_{j_1}...z_{j_p}$;
the irreducible representations are thus of the tensorial
type $T^q_p$ with $p$ symmetric lower indices, $q$ symmetric 
upper indices and
the contraction (or trace) of any $p$-type 
index with any $q$-type index must vanish.
 On these, $\L$ has the value $il$,
$l=p-q$. 
Each such representation contains a unique $SU(k-1)$-invariant state,
which is of the form $\vert 0,p,q\ra \sim \bz_k^q z_k^p$.
Comparing this with $z_i = h_{i,k^2-1}$ we see that
$l$ may be identified as the eigenvalue of $t_{k^2-1}$ 
of $SU(k)$ acting on the right of $h$. The mode expansion 
of the field $\Phi$ may therefore be written as
\beq
\Phi = \sum_l \sum_{p,q|p-q=l}~c^{p,q}_m ~\D^{(p,q)}_{m;(0,p,q)} (h)
\label{44}
\eeq
Notice that $\D^{(p,q)}_{m;(0,p,q)}$ are sections of $U(1)$-bundles
on ${\bf CP}^{k-1}$ and are therefore similar to wavefunctions
of a reduced Landau problem on ${\bf CP}^{k-1}$ with
different choices of magnetic field.

\section{${\bf CP}^1$ case} 

We now illustrate briefly how our approach works for the simple case
of quantum Hall effect on ${\bf{CP}}^1= S^2 = {{SU(2)} / U(1)}$. 

In this case the $SU(2)$ group element $g$ can be 
parametrized in terms of the complex variables $u_1,~u_2$,
where $u^*\cdot u =1$
\beq
g = \left( \matrix {u_2^* & u_1 \cr
-u_1^* & u_2} \right)
\label{441}
\eeq
and
\beq
u = {1 \over {\sqrt{1 +\xi \bxi}}} \left( \matrix{\xi \cr
1\cr} \right)
\label{442}
\eeq
where $\xi,~\bxi$ are the local complex coordinates for $S^2$.

Given (\ref{441}), (\ref{442}) one can derive expressions for both the
left and right generators in terms of $u$'s. In particular
\beqar
\hat{R}_{+} = -\ep _{\alpha \beta} u^*_{\alpha } {\del \over {\del
    u_{\beta}}} = (1 + \xi \bxi) { \del \over {\del \xi}} \nonumber \\
\hat{R}_{-} = \ep _{\alpha \beta} u_{\alpha} {\del \over {\del
    u_{\beta}^*}} = -(1 + \xi \bxi) { \del \over {\del \bxi}} 
\label{443}
\eeqar
Using (\ref{443}) we find
\beq
([A,B]) (g) = -{1 \over n} (1 + \xi \bxi)^2 \left( \del_\xi A \del_{\bxi} B
- \del_{\bxi} A \del_\xi B \right)
\label{444}
\eeq
Comparing this with (\ref{31}) we easily verify that (\ref{30}) holds.

The $U(1)$ invariant confining potential that allows the formation of a droplet
with density of the form (\ref{34}) is
\beq
\hat{V} = \omega ( \hat{L} _3 + {n \over 2})
\label{445}
\eeq
with the corresponding function
\beq
V(\xi, \bxi) = \omega  n {{\xi \bxi} \over {1 + \xi \bxi}}
\label{446}
\eeq
This is essentially a harmonic oscillator potential for large
$n$. Further
\beqar
\{ \Phi,~ \rho_0 \} &=& i ( 1 + \xi \bxi)^2  \left( \del_\xi \Phi
\del_{\bxi} 
\rho_0
- \del_{\bxi} \Phi \del_\xi \rho_0 \right) \nonumber \\
& = & i ( 1 + \xi \bxi)^2  \left( \xi
\del_{\xi} 
-  \bxi \del_{\bxi}\right) \Phi { {\del \rho_0} \over {\del r^2}}
\label{447}
\eeqar
which identifies the operator in (\ref{12c}) as 
$\L = \del _{\theta}$, where $\xi = r e^{i \theta}$.

Using (\ref{446}), (\ref{447}) in (\ref{12}) we find the well known
  one-dimensional chiral bosonic action describing edge excitations for
  $\nu =1$ two-dimensional QHE \cite{stone}
\beq
S = -{1 \over {4 \pi}} \oint d\theta ~  
\left( \del_t +
 \omega \del_{\theta}  \right) \Phi (\theta , t)~\del_{\theta} \Phi(\theta, t)
\label{448}
\eeq

\section{Edge states for $S^4$ from ${\bf CP}^3$}

As mentioned in section 3, even for the case of a $U(1)$
background magnetic field on ${\bf CP}^k$, one can get
somewhat different results depending on the choice of
the potential and how the states are filled to
form the droplet. Among these possibilities, the case
of ${\bf CP}^3$ is especially interesting because
of the fact that it is an $S^2$ bundle over $S^4$.
Edge states for the Hall effect on $S^4$ with an
instanton background field can be obtained
using a $U(1)$ background field on ${\bf CP}^3$.
Here we will obtain the edge effective action by reducing
our general action for the case of ${\bf CP}^3$ to
$S^4$.

The space ${\bf CP}^3$ is realized as $SU(4)/U(3)$, so
that one can use the group elements as local coordinates,
as we have done so far. But one can also use the homogeneous
coordinates and this is more convenient for the reduction to $S^4$.
We can describe ${\bf CP}^3$ by the four complex coordinates
$Z_i$, $i=1,...,4$, with the identification
$Z_i \sim \lambda Z_i$ where $\lambda$ is any complex number
except zero, $\lambda \in {\bf C}-\{ 0 \}$.
We write $Z_i$ in terms of two two-component
spinors $w ,~\pi$ as
\beq
(Z_1, Z_2, Z_3, Z_4)= (w_1, w_2, \pi_1, \pi_2)
\label{45}
\eeq
Coordinates $x_\mu$ on $S^4$ are then defined by
\beq
w = (x_4 -i \sigma \cdot x ) ~\pi
\label{rev3}
\eeq
The scale invariance $Z\sim \lambda Z$ can be realized
as the scale invariance $\pi \sim \lambda \pi$;
the $\pi$'s then describe a ${\bf CP}^1 =S^2$. This will be
the fiber space. The coordinates $x_\mu$ are the standard
stereographic coordinates for $S^4$; one can in fact write
\beq
y_0 = {1-x^2 \over 1+x^2},\hskip 1in y_\mu = {2x_\mu \over 1+x^2},
\label{46}
\eeq 
to realize the $S^4$ as embedded in ${\bf R}^5$.
The definition of $x_\mu$ in terms of $w$ may be
solved as
\beqar
x_4 &=& {1\over 2} {{\bar \pi}w +{\bar w}\pi \over
{\bar \pi}\pi}\nonumber\\
x_i&=& {i\over 2} {{\bar \pi} \sigma_i w - {\bar w}
\sigma_i \pi \over {\bar \pi} \pi}
\label{47}
\eeqar

The K\"ahler two-form on ${\bf CP}^3$ is given as
\beq
\Omega = - i \left[ {d{\bar Z}\cdot dZ \over {\bar Z}\cdot Z}
- {d {\bar Z}\cdot Z ~{\bar Z}\cdot dZ \over ({\bar Z}\cdot Z)^2}
\right] \label{48}
\eeq
Notice that this is invariant under $Z\rightarrow \lambda Z$,
and ${\bar Z}\rightarrow \lambda {\bar Z}$.
We can reduce this using (\ref{45}), (\ref{47})
to get
\beqar
\Omega_{{\bf CP}^3} &=& \Omega_{{\bf CP}^1} - i~F\nonumber\\
F&=& dA +A ~A\nonumber\\
A&=& i {N^a \eta^a _{\mu\nu} x^\mu dx^\nu \over (1+x^2)}
\label{49}
\eeqar
Here $\eta^a_{\mu\nu}$
is the 't Hooft tensor
\beq
\eta^a_{\mu\nu}= \ep_{a\mu\nu 4} + \delta_{a\mu} \delta_{4\nu}
- \delta_{a\nu} \delta_{4\mu} \label{50}
\eeq
and
\beq
N^a= {\bar \pi} \sigma^a \pi /{\bar \pi}\pi\label{50a}
\eeq
$N^a$ is a unit
three-vector, which may be taken as parametrizing the fiber
${\bf CP}^1 \sim S^2$.
The field $F$ is the instanton field. We see that we
can get an instanton background on $S^4$ by taking
a $U(1)$ field on ${\bf CP}^3$ proportional to the
K\"ahler form.

Functions $A$ on ${\bf CP}^3$ can be considered as functions
of the four variables $Z_i$, but because of the
homogeneity conditions, they must obey
\beq
Z\cdot {\del A \over \del Z}=0,\hskip 1in
{\bar Z}\cdot {\del A \over \del {\bar Z}} =0
\label{51}
\eeq
The K\"ahler form can be inverted on functions obeying this
condition to get the Poisson bracket as
\beq
\{ A, B \} =
i ({\bar Z}\cdot Z)~ \left( {\del A \over \del {Z}_\alpha}
{\del B \over \del \bZ_\alpha} - {\del A \over \del \bZ_\alpha}
{\del B \over \del {Z}_\alpha}\right)
\label{52}
\eeq
It is now straightforward to simplify this in terms
of the $S^4$ coordinates, using (\ref{45}), (\ref{47})
to get
\beqar
\{ A, B\}_{{\bf CP}^3}
&=& (1+x^2) K^{\mu\nu}{\del A \over \del x^\mu}
{\del B\over \del x^\nu} ~+~ (1+x^2) \{ A, B\}_{{\bf CP}^1}
\nonumber\\
K_{\mu\nu}&=& - {1\over 2} N^a \eta^a_{\mu\nu}
\label{53}
\eeqar
The second term is the Poisson bracket for ${\bf CP}^1$.
$K_{\mu\nu}$
defines a local complex structure,
a way of combining
the $x_\mu$ into complex coordinates; for every point on
$S^2$ given by $N^a$ we have a corresponding $K_{\mu\nu}$.
This is in accordance with ${\bf CP}^3$ being a bundle
of local complex structures on $S^4$.

The Poisson bracket, split as in (\ref{53}), is one of the ingredients
for the effective action.
We will now turn to the density matrix.
The required density should depend only on the
coordinates of $S^4$, since we want to interpret this as
a droplet in $S^4$. We take it to be of the form
\beq
{ \rho}_0 = \sum_{s=0}^M {n! \over s! (n-s)!}
{(x^2)^s \over (1+x^2)^n}
\label{54}
\eeq
This will behave like a step function of radius $\sim\sqrt{M}$ 
and so corresponds to a droplet on $S^4$. In terms of
the homogeneous coordinates $Z=(w, \pi)$ we find that
this is
\beq
{\rho}_0 = \sum^M {n! \over s! (n-s)!}
{{(\bZ_3 Z _3 + \bZ_4 Z_4)^{n-s} (\bZ_1 Z_1 +\bZ_2 Z_2)^s} \over 
{(\bZ \cdot Z)^n}}
\label{55}
\eeq
By expanding out the numerator, identifying 
$\xi_i = {Z_i /Z_4},~i=1,2,3$,
and comparing with the coherent states
defined earlier in (\ref{17}), 
we can identify the filled states.
We define an $SU(2)$ subgroup of $SU(4)$ as the
group acting on $\xi_1, \xi_2$.
The $SU(4)$
representation of the lowest
Landau level can be split into
multiplets of this $SU(2)$, which we call the
right isospin with
$I_R = \half s$, where $s$ is the number
of $\xi_1$ or $\xi_2$ as in (\ref{55}).
(The right isospin will correspond to $SU(2)_R$
in a decomposition of the $O(4)$ algebra of $S^4$ as
$O(4) \sim SU(2)_L \times SU(2)_R$.)
The density (\ref{54}) is produced 
when all the $SU(2)_R$ isospin multiplets up to $s=M$ are filled. 
The potential which will lead to filling the various
isospin multiplets, from $s=0$ to $s=M$, can be taken as
proportional to $s$ itself, namely 
\beq
\la s | \hat{V} | s \ra = \omega s
\label{55a}
\eeq
Its expression in terms of Lie algebra generators is
\beq
\hat{V} =  {\omega\over 6} (t_{15} + 2 t_8 +3n )
\label{56}
\eeq
The calculation of the corresponding classical function as in 
(\ref{38})-(\ref{38c}) gives
\beq
V = \omega n {{\bxi_1 \xi_1 + \bxi_2 \xi_2} \over {1 + \bxi \cdot
    \xi}} = 
\omega n
{x^2 \over {1 + x^2}}
\label {56a}
\eeq
With this potential, we then get the density 
in (\ref{54}).

The density  does not depend on the ${\bf CP}^1$ coordinates.
As a result, the contribution from the ${\bf CP}^1$ Poisson brackets
will integrate to zero in the action.
The action then becomes
\beq
S = - {M\over 4\pi^2} ~n\int d\mu_{{\bf CP}^1}~
\int d\Omega_3 \left[ {\del\Phi \over \del t} (\L\Phi ) +
\omega (\L \Phi )^2 \right]\label{58}
\eeq
where $\L\Phi = 2x^\nu K^{\mu\nu} \del_\mu \Phi$. $\Phi$'s are to
be expanded
in terms of harmonics on ${\bf CP}^1$ which correspond to
the representations of $SU(2)_L$ in $O(4) \sim
SU(2)_L\times SU(2)_R$. (This is the subgroup
corresponding to the instanton gauge group.) 
Recall that the
inner product for states on ${\bf CP}^1$
has a factor which is the dimension of the representation in it, 
and so we must interpret
the prefactor $n$ for the integral over $d\mu_{{\bf CP}^1}$
as this quantity. As $n\rightarrow \infty$, this becomes
very large; thus the number of
left isospin states becomes very large, in agreement
with the analysis of
\cite{HZ}.

We will conclude this section with a few remarks on the nature of the excitations
in (\ref{58}). The equation of motion for $\Phi$ is of the form
$\L ~\left( i \del_t \Phi + i \omega \L \Phi \right) =0$.
The zero momentum state with $\L \Phi =0$ does not give
a deformation of the boundary and so, for states of interest, the equation of motion implies
\beq
 i \del_t \Phi = - i \omega \L \Phi  
\label{add1}
\eeq
To understand the nature of the states, we need to seek eigenstates
of $\L$. By direct application on the $w, ~{\bar w}$ in (\ref{rev3}),
we see that
\beqar
- i \L ~w_{\dot A} &=& ~w_{\dot A}\nonumber\\
- i \L~ {\tilde w}_{\dot A} &=&- {\tilde w}_{\dot A}
\label{add2}
\eeqar
where ${\tilde w}_{\dot A} = \epsilon_{{\dot A} {\dot B}} {\bar w}_{\dot B}$.
Eigenstates of $\L$ are thus given by a monomial with a certain number of
$w$'s and ${\tilde w}$'s and an arbitrary number of $\pi$'s and  $\tilde{\pi}$'s, where
$\tilde{\pi}_{C}= \epsilon_{CD} \bar{\pi}_{D}$
\beqar
&&-i\L ~ {\tilde w}_{\dot A_1}\cdots {\tilde w}_{\dot A_m}
w_{\dot B_1}  \cdots w_{\dot B_l} \tilde{\pi}_{C_1}  \cdots
\tilde{\pi}_{C_q} \pi_{D_1}  \cdots \pi_{D_k} \nonumber \\
&&= (l-m) ~{\tilde w}_{\dot
A_1}\cdots {\tilde w}_{\dot A_m} w_{\dot B_1} \cdots w_{\dot
B_l} \tilde{\pi}_{C_1}  \cdots
\tilde{\pi}_{C_q} \pi_{D_1} \cdots \pi_{D_k}
\label{add3}
\eeqar
Since $\epsilon_{{\dot A} {\dot B}}{\tilde w}_{\dot A}
w_{\dot B} = x^2 {\bar \pi }\cdot \pi $ and $x^2$ is fixed on the boundary of the droplet, such
contractions may be removed from (\ref{add3}). One may consider all the $\{\dot{A},
\dot{B} \}$ indices to be symmetric and similarly all the $\{ C, D \}$ indices to be
symmetric, making the eigenstates of $\cal{L}$ irreducible tensors of $SU(2)_R \times
SU(2)_L$. Further, since $\Phi$ is a function on
${\bf CP}^1$, we must have invariance under the scaling $\pi \rightarrow
\lambda \pi$. The mode expansion for $\Phi$ in terms of eigenstates of
$\L$ is given by
\beqar
\Phi &=& \sum_{l\geq m} ~ 
C^{(\dot A)_m (\dot B)_l (C)_{l-m+k} (D)_k} ~f_{(\dot A)_m (\dot B)_l (C)_{l-m+k}
(D)_k}\nonumber\\ &&+\sum_{l < m} ~ 
{\tilde C}^{(\dot A)_m (\dot B)_l (C)_{k} (D)_{m-l+k}}~{\tilde f}_{(\dot A)_m (\dot B)_l
(C)_{k} (D)_{m-l+k}}\nonumber\\ f_{(\dot A)_m (\dot B)_l (C)_{l-m+k}
(D)_k}&=&
{1\over ({\bar \pi}\cdot \pi )^{l+k}}~{\tilde w}_{\dot A_1}
\cdots  {\tilde w}_{\dot A_m}
w_{\dot B_1} \cdots w_{\dot B_l} {\tilde \pi}_{C_1} \cdots {\tilde \pi}_{C_{l-m+k}}
\pi_{D_1} \cdots \pi_{D_k} \nonumber\\ {\tilde f}_{(\dot A)_m (\dot B)_l
(C)_{k} (D)_{m-l+k}}
 &=& {1\over ({\bar \pi}\cdot \pi )^{m+k}}~{\tilde w}_{\dot A_1}
\cdots  {\tilde w}_{\dot A_m}
w_{\dot B_1} \cdots w_{\dot B_l} {\tilde \pi}_{C_1} \cdots {\tilde \pi}_{C_k} \pi_{D_1} \cdots
\pi_{D_{m-l+k}} 
\label{add4}
\eeqar
where $({\dot A})_m = {\dot A}_1 \cdots {\dot A}_{m}$, $(C)_k = C_1 \cdots C_k$ and similarly
for the other indices. Each function $f$ ($\tilde{f})$ transforms as an
irreducible representation of $SU(2)_L \times SU(2)_R$, with
the $j$-values $\half \vert l-m\vert + k$ and $\half (l+m)$ respectively. 

In comparing with the equation of motion, (\ref{add1}), we see that all these states satisfy
the dispersion relation $E= -i \omega {\cal{L}} \sim P_3$, where $P_3$ is the linear momentum
corresponding to $-i \L$. All states with the same value for
$-i\cal{L}$, namely fixed $\vert m-l \vert$, are degenerate in terms of energy. Out of these,
there is a subset of states, which satisfy the relativistic dispersion relation $E \sim P_3=
\sqrt{P^2}$. These are the states with $P_1=P_2=0$, where $P_1,~P_2$ are the other two
directions of momentum on the boundary of the droplet. The three momentum operators do not
commute with each other.

From (\ref{add2}) the operator $-i\cal{L}$ can be written as
\beq
-i {\cal{L}} = w_{\dot A} {\del \over \del w_{\dot A}} - {\tilde w}_{\dot A} {\del \over \del
{\tilde w}_{\dot A}}
\label{rev1}
\eeq
We now define the 
operators
\beqar
\L_+ &=& w_{\dot A} {\del \over \del {\tilde w}_{\dot A}}\nonumber\\ 
\L_- &=& {\tilde w}_{\dot A} {\del \over \del w_{\dot A}}
\label{add5}
\eeqar
These form an $SU(2)$ algebra along with
$-i \L$ and correspond to translations along the two other directions (1 and 2) on the
boundary of the droplet. Among the modes in (\ref{add4}), there are special states, the
highest (lowest) weight states which obey the condition
\beq
\L_+ f  =0~~~~~{\rm or}~~~~~ \L_- \tilde{f} =0
\label{rev2}
\eeq
They are of the form
\beqar
f_{{\dot B_1}
 \cdots {\dot B_l} { C_1}\cdots { C_{l+k}} {D_1} \cdots {D_k}}&=&
{1\over ({\bar \pi}\cdot \pi )^{l+k}}~
w_{\dot B_1} \cdots w_{\dot B_l} {\bar \pi}_{C_1} \cdots {\bar \pi}_{C_{l+k}}
\pi_{D_1} \cdots \pi_{D_k} \nonumber\\ {\tilde f}_{{\dot A_1} \cdots {\dot A_l}
{ C_1}\cdots { C_k} {D_1} \cdots {D_{l+k}}}
 &=& {1\over ({\bar \pi}\cdot \pi )^{l+k}}~{\tilde w}_{\dot A_1}
\cdots  {\tilde w}_{\dot A_l}
{\tilde \pi}_{C_1} \cdots {\tilde \pi}_{C_k} \pi_{D_1} \cdots \pi_{D_{l+k}}
\label{add6}
\eeqar
For these states, the expectation value of $P_1^2 + P_2^2 = (\L_- \L_+ +\L_+
\L_-)/r_D^2$ tends to zero as the size of the droplet becomes very large, compared to $P_3$
which remains finite. Their momentum is
essentially all given by $P_3 = -i\L/r_D$.
These states therefore satisfy the relativistic
dispersion relation $ E = -i \omega \L \sim P_3 \sim \sqrt{P^2}$.
(These are related to the
extremal dipole states in \cite{HZ}, \cite{pol}.) For states which are not of highest (or
lowest) weight,
$E
\sim P_3$ but this is no longer the square root of
$P^2$ since
$P_1^2,~P_2^2$ are not zero. 

Since the modes transform as irreducible representations of
$SU(2)_L \times SU(2)_R$, a consistent definition of helicity would be
the difference of the $j$-values for the two $SU(2)$'s. For
the highest (lowest) weight states, which satisfy the relativistic dispersion relation, this
value is $k$.   These results are in general agreement with the 
results on edge states in \cite{{HZ}, {pol}}. 

We have found that the effective edge action describing the particle-hole like
excitations of the Zhang-Hu four dimensional quantum Hall droplet does not describe a
relativistic theory. Whether and how one can consistently truncate the spectrum to keep only
the highest (lowest) weight states in order to obtain a relativistic theory is not clear. 

\section{Conclusions}

We have presented a method for obtaining
the general form of the effective action
for edge excitations of a quantum Hall droplet in higher
dimensions. We analyzed in detail the case of $\nu =1$ QHE on even
dimensional ${\bf {CP}}^k$ spaces with a $U(1)$ background magnetic
field, which admit droplet configurations in the presence of a
confining potential. We find that the edge excitations can be
described, in the limit of large number of fermions, in terms of
a chiral action for an abelian bosonic field. The chirality here is 
expressed by the fact that there is a preferred direction along the
boundary of the droplet which is relevant for the dynamics of the edge
excitations. This tangential direction is the Poisson conjugate to the
normal to the droplet, with the Poisson structure given by the 
K\"ahler
form of the underlying space where the fermions live, thus it is
determined by the geometry of the underlying space. 
We have also obtained, starting from ${\bf CP}^3$, the
edge action for QHE droplets on $S^4$, the original model of
Zhang and Hu \cite{HZ}. We have clarified the kind of potential
needed and the corresponding density function for
the reduction to $S^4$ to go through. The corresponding effective edge action does not
describe a relativistic theory, although there are energy eigenmodes in the decomposition of
the bosonic field $\Phi$ that satisfy the relativistic dispersion relation.

We emphasize that there
is a general pattern to all these cases.
For ${\bf CP}^k$, the K\"ahler structure determines
the form of the action. $S^4$ does not have a complex
structure, but ${\bf CP}^3$ can be viewed as the bundle of local
complex structures on $S^4$; this is the twistor
space approach. And not surprisingly,
the edge action for $S^4$ is determined
by the local complex structure.
In reference \cite{berneveg}, an effective description of the
Hall droplet in terms of a Chern-Simons theory is obtained. It would be
interesting to see if our action for edge excitations can be related
to this bulk theory via gauge anomaly arguments.

The analysis in this paper has been for abelian background
magnetic fields and for one species of fermions.
If we have a nonabelian background field, or if we have more 
than one species of fermions, the edge excitations are described by
an element of a nonabelian group. The resulting edge action
will be related to a chiral Wess-Zumino-Witten theory.
This will be described in more detail elsewhere
\cite{KN3}.

\vskip .1in
\noindent{\bf Acknowledgements}

We thank the Abdus Salam International Centre for Theoretical
Physics for hospitality during the course of this work.
This work was supported in part by the National Science Foundation
under grant numbers PHY-0140262 and PHY-0244873 and
by PSC-CUNY grants.

\end{document}